\documentclass[aps,pra,twocolumn,superscriptaddress,showpacs,showkeys,amsmath,amssymb]{revtex4}

\usepackage{amsfonts}
\usepackage{amssymb,amsmath}
\usepackage{mathrsfs}
\usepackage{latexsym}
\usepackage{amsmath}
\usepackage[cp1251]{inputenc}
\usepackage{graphicx}
\usepackage{dcolumn}
\usepackage{bm}
\usepackage{color}

\begin{document}

\title{Infrared behavior of dipolar Bose systems at low temperatures}

\author{Volodymyr~Pastukhov\footnote{e-mail: volodyapastukhov@gmail.com}}
\affiliation{Department for Theoretical Physics, Ivan Franko National University of Lviv,\\ 12 Drahomanov Street, Lviv-5, 79005, Ukraine}

\date{\today}



\begin{abstract}
We rigorously discuss the infrared behavior of the uniform three
dimensional dipolar Bose systems. In particular, it is shown that
low-temperature physics of the system is controlled by two
parameters, namely isothermal compressibility and intensity of the
dipole-dipole interaction. By using a hydrodynamic approach we
calculate the spectrum and damping of low-lying excitations and
analyze the infrared behavior of the one-particle Green's function.
The low-temperature corrections to the anisotropic superfluid
density as well as condensate depletion are found. Additionally we
derive equations of the two-fluid hydrodynamics for dipolar Bose
systems and calculate velocities of first and second sound.
\end{abstract}

\maketitle

\section{Introduction}
\label{sec1}
\setcounter{equation}{0}

The experimental realization of the Bose-Einstein condensation of
atoms with large magnetic moments such as chromium ({$^{52}$Cr})
\cite{Griesmaier} and more recently dysprosium ({$^{164}$Dy})
\cite{Lu} and erbium ($^{168}$Er) \cite{Aikawa} stimulated
extensive theoretical studies of these systems
\cite{Baranov,Lahaye_Menotti,Baranov_Dalmonte}. The presence of
the comparatively strong dipole-dipole interaction gives rise to
exciting properties of the dipolar Bose condensates, namely the
direction-dependent spectrum of elementary excitations,
anisotropic superfluidity \cite{Wilson_Ronen} at finite
temperatures, non-typical soliton formation
\cite{Pedri,Tikhonenkov}, and two types of the ground state
\cite{Bisset_Wilson}. In some configurations of the trapping
potential the roton character \cite{ODell,Santos} of the spectrum
of collective modes is intrinsic for dipolar condensates providing
the remarkable phenomena like spatial roton confinement
\cite{Jona-Lasinio_Lakomy_Santos} and anomalous atom-number
fluctuations \cite{Bisset_Blakie_13}.

Despite the trapped case where even a fully dipolar Bose gas can be
realized by using a strong harmonic potential that confines the
system in the plane perpendicular to external magnetic field
\cite{Koch} for the stabilization of three-dimensional homogeneous
condensates with dipole-dipole interaction the short-range
repulsion between particles is required. Fortunately, the strength
of this repulsive term can be tuned by Feshbach resonance
techniques allowing to study the collapse dynamics and expansion
of dipolar condensates \cite{Lahaye_Metz}. It was usually believed
\cite{Goral,Eberlein} that the properties of these systems can be
understood on the mean-field level but the recent observation of droplet formation in dysprosium condensates \cite{Kadau}
breaks this notion providing the influential role of quantum
fluctuations \cite{Schutzhold,Lima_Pelster, Wachtler_Santos}. Recent
path-integral Monte Carlo simulations \cite{Saito} confirm these
findings. It is interesting that the mean-field impact of three-body
repulsion forces also leads to stable droplets
\cite{Bisset_Blakie,Xi_Saito} in dipolar Bose condensates. At
finite temperatures the calculations with dipolar condensates are
complicated but still tractable \cite{Bisset_Baillie_Blakie},
where the role of the dipole-dipole interaction and trap geometry
on the thermodynamics of dipolar Bose gases was discussed.

The purpose of the present paper is to explore exact low-energy and,
in turn, low-temperature properties of dipolar Bose systems.

\section{Formulation of the model}
\label{sec2} \setcounter{equation}{0}
We consider a system of $N$ spinless particles immersed in volume $V$ with the
second-quantized Hamiltonian
\begin{eqnarray}\label{H}
H=H_0+\Phi,
\end{eqnarray}
were the first term is the kinetic energy operator
\begin{eqnarray}
H_0=-\frac{\hbar^2}{2m}\int d{\bf r}\psi^+({\bf r}){\bf \nabla}^2
\psi( {\bf r})
\end{eqnarray}
and $\Phi$ takes into account the pairwise interaction between
particles
\begin{eqnarray}
\Phi=\frac{1}{2}\int d{\bf r}\int d{\bf r}'\Phi({\bf r}-{\bf r}')
\psi^+( {\bf r})\psi^+( {\bf r}')\psi({\bf r}')\psi({\bf r}).
\end{eqnarray}
The field operators satisfy the usual bosonic commutation relations
$[\psi({\bf r}), \psi^+({\bf r}')]=\delta({\bf r}-{\bf r}')$,
$[\psi({\bf r}), \psi({\bf r}')]=0$. The two-particle potential
$\Phi({\bf r})$ necessarily contains dipole-dipole interaction
\begin{eqnarray}\label{Phi_d}
\Phi_{d}({\bf
r})=\frac{3g_{d}}{4\pi}\left\{\frac{1}{r^3}-\frac{3z^2}{r^5}\right\},
\end{eqnarray}
and any additional term to stabilize the system. Actually, the
results obtained below are not dependent on the specific form of
this part of the two-body interaction, except it must be repulsive
enough at small inter-particle separations.

Avoiding the problem with infrared divergences
\cite{Nepomnyashchy} we adopt the scheme normally used for
low-dimensional systems \cite{Mora_Castin}
\begin{eqnarray}\label{psi}
\psi({\bf r})=e^{i\varphi({\bf r})}\sqrt{\rho({\bf r})}, \ \
\psi^+({\bf r})=\sqrt{\rho({\bf r})}e^{-i\varphi({\bf r})},
\end{eqnarray}
where we introduce phase $\varphi({\bf r})$ and density operators
$\rho({\bf r})=\psi^+( {\bf r})\psi({\bf r})$ with commutator
$[\rho({\bf r}'),\varphi({\bf r})]=i\delta({\bf r}-{\bf r}')$.
After this substitution the Hamiltonian of the dipolar Bose gas
reads
\begin{eqnarray}\label{H_rho_phi}
H[\varphi({\bf r}), \rho({\bf r})]=\frac{1}{2}\int d{\bf r}\int d{\bf
r}'\Phi({\bf r}-{\bf
    r}') \rho( {\bf r})\rho( {\bf r}') \nonumber\\
+\frac{\hbar^2}{2m}\int d{\bf r}\left\{ \frac{( \nabla \rho({\bf
r}))^2}{4 \rho({\bf r})} +\rho({\bf r})(\nabla \varphi({\bf r}))^2
\right\},
\end{eqnarray}
where the inconsequential constant term that shifts the ground-state
energy of the system is omitted. Actually, Eq.~(\ref{H_rho_phi}) is
the Hamiltonian obtained for the first time in
Ref.~\cite{Bogoliubov_Zubarev} and written here explicitly in the
Hermitian form. In order to derive the original Bogoliubov-Zubarev
result we have to use the following representation of field
operators: $\psi({\bf r})\sim e^{i\varphi({\bf r})}$, $\psi^+({\bf
r})\sim \rho({\bf r})e^{-i\varphi({\bf r})}$. In context of liquid
$^4$He theory the Hamiltonian (\ref{H_rho_phi}) was studied
extensively in Refs.~\cite{herm_ham}.

To proceed we pass to the path-integral representation
\cite{Popov}. In doing so we have to introduce new fields with
explicit imaginary time $\tau \in [0, \beta]$ dependence (where
$\beta=1/T$ is the inverse temperature of the system). Then the
partition function reads
\begin{eqnarray}\label{Z}
Z=\int D\varphi D\rho e^{S},
\end{eqnarray}
where the functional integral is carried out over $\beta$-periodic
fields $\varphi(x)$, $\rho(x)$ ($x=(\tau, {\bf r})$) with action
\begin{eqnarray}\label{S_hydro}
S=\int^{\beta}_0 d\tau\int d{\bf
r}\{\rho(x)i\partial_{\tau}\varphi(x)+\mu \rho(x)\}\nonumber\\
-\int^{\beta}_0 d\tau H[\varphi(x), \rho(x)].
\end{eqnarray}
Here we made the following replacement in the Hamiltonian
$H[\varphi({\bf r}), \rho({\bf r})]\rightarrow H[\varphi(x),
\rho(x)]$ and introduced the chemical potential $\mu$ of the
system. For the spatially uniform system we can use the following
decomposition of the density and phase fields
\begin{eqnarray}\label{n_a}
\rho(x)=\rho(\tau)+\frac{1}{\sqrt{V}}\sum_{{\bf k}\neq 0}e^{i{\bf
kr}}\rho_{\bf k}(\tau), \nonumber\\
\varphi(x)=\varphi(\tau)+\frac{1}{\sqrt{V}}\sum_{{\bf k}\neq
0}e^{i{\bf kr}}\varphi_{\bf k}(\tau).
\end{eqnarray}
From the structure of action (\ref{S_hydro}) it is clearly seen
that the only term containing zero-momentum $\varphi$-variables is
$V\int^\beta_0d\tau \rho(\tau)i\partial_{\tau}\varphi(\tau)$.
The integration over $\varphi(\tau)$ results in delta-functions that
require $V\rho(\tau)$ to be a constant (independent of $\tau$)
integer positive number \cite{Wilson_Galitski}, so for the
partition function we have
\begin{eqnarray}
Z=\sum_{\mathscr{N}\geq 0}\exp\left\{\beta V[\mu
\mathscr{N}/V-f(\mathscr{N}/V)]\right\},
\end{eqnarray}
where $e^{-\beta Vf( \mathscr{N}/V)}$ denotes the part of
Eq.~(\ref{Z}) with $\rho_{\bf k}(\tau)$,
$\varphi_{\bf k}(\tau)$ integrated out and $\rho(\tau)$ replaced by
$\mathscr{N}/V$. In the thermodynamic limit when the number of
particles $N$ together with volume $V$ of the system tend to
infinity (but $N/V={\rm const}=\rho$), the summation in the above
formula can be easily changed by integration, which we perform
using the steepest descent method. Finally, the asymptotically exact
expression for the partition function is
\begin{eqnarray}
Z=\exp\{\beta V[\mu \rho-f(\rho)]\}, \ \ \mu=\left(\frac{\partial
f}{\partial \rho}\right)_{T},
\end{eqnarray}
and therefore we can proceed our consideration in the canonical
ensemble.

\section{Low-temperature behavior}
\label{sec2}
\subsection{Low-energy excitations}
In the previous section it was argued that considering only part
of action (\ref{S_hydro}) with non-zero momentum modes we can
calculate the free energy of the system $Vf(\rho)$ identifying
$\rho$ as the equilibrium (uniform) density. Now we are in position to
formulate the theory in terms of density and phase fluctuations.
The corresponding action after series expansion of
$\rho(x)$-fields near $\rho$ reads
\begin{eqnarray}\label{S}
S=S_{0}+S_{int},
\end{eqnarray}
where $S_0$ is the Gaussian term
\begin{eqnarray}\label{S_0}
S_{0}={\rm{const}}-\frac{1}{2}\sum_{K}\left\{\omega_{k}\varphi_K\rho_{-K}
-\omega_{k}\varphi_{-K}\rho_{K}
\right.\nonumber\\
\left.+\frac{\hbar^2k^2}{m}\rho\varphi_{K}\varphi_{-K}+\left[\frac{\hbar^2k^2}{4m\rho}+\nu({\bf
k})\right]\rho_{K}\rho_{-K}\right\},
\end{eqnarray}
here we introduce the notation for the four-momentum $K=(\omega_k,
{\bf k})$, where $\omega_k$ is the bosonic Matsubara frequency;
$\nu({\bf k})$ is the Fourier transform of $\Phi(\bf{r})$;
${\rm{const}}=-\beta V\rho^2\nu(0)/2+\beta\frac{1}{2}\sum_{{\bf
k}\neq 0}[\rho\nu({\bf k})+\hbar^2 k^2/2 m ]$ and $\nu(0)$ should
be treated as a direction-averaged value of $\nu({\bf k})$ in the
${\bf k} \rightarrow {\bf 0}$ limit (see discussion in
Ref.~\cite{Yukalov_Yukalova}). Thus, due to the homogeneity of the
system the dipolar interaction does not affect on the mean-field
ground-state energy. On the other hand, the presence of external
non-uniform potential that breaks continuous translation
invariance and causes a small deviation $\delta\rho({\bf r})$ in the
density distribution shifts the thermodynamic potentials by $\frac{1}{2}\int
d{\bf r}\int d{\bf r}'\delta\rho({\bf r}')\Phi_{d}({\bf r}'-{\bf
r})\delta\rho({\bf r})$.

The last term in Eq.~(\ref{S}) takes into account the simplest
collision processes between elementary excitations
\begin{eqnarray}\label{S_int}
S_{int}&=&-\frac{1}{2\sqrt{\beta V}}\sum_{K,
Q}D^{(0)}_{\varphi\varphi\rho}(K,Q|-K-Q)
\varphi_{K}\varphi_{Q}\rho_{-K-Q}\nonumber\\
&-&\frac{1}{3!\sqrt{\beta
V}}\sum_{K+Q+P=0}D^{(0)}_{\rho\rho\rho}(K,Q,P)\rho_{K}\rho_{Q}\rho_{P},
\end{eqnarray}
where bare vertices
$D^{(0)}_{\varphi\varphi\rho}(K,Q|-K-Q)=-\frac{\hbar^2}{m}{\bf
kq}$ and
$D^{(0)}_{\rho\rho\rho}(K,Q,P)=-\frac{\hbar^2}{8m\rho^2}(k^2+q^2+p^2)$.
In fact, $S_{int}$ contains an infinite series, but we present only
the relevant terms for our low-temperature description.

Recently \cite{Pastukhov_q2D} we showed how to relate the infrared
asymptotics of the matrix correlation function
\begin{eqnarray}\label{D_K}
{\bf D}^{-1}(K)=\left( \begin{array}{c c}
\langle\varphi_K\varphi_{-K}\rangle & \langle\rho_K\varphi_{-K}\rangle\\
\langle\varphi_K\rho_{-K}\rangle & \langle\rho_K\rho_{-K}\rangle\\
\end{array} \right),
\end{eqnarray}
with macroscopic parameters of the system. For later convenience
we introduce the notation for the second-order vertices ${\bf
D}(K)={\bf D}_0(K)-{\bf \Pi}(K)$, where ${\bf \Pi}(K)$ is the
self-energy matrix and ${\bf D}_0(K)$ is given by Eq.~(\ref{S_0}).
Particularly, it was demonstrated that the sound velocity of
low-lying excitations can be uniquely written in terms of the
superfluid density and inverse susceptibility. It should be noted
that the same behavior of the low-energy spin-wave excitations is
observed in the Heisenberg and easy-plane antiferromagnets
\cite{Halperin_Hohenberg,Chernyshev_Zhitomirsky}. This analysis
can be naturally extended on the Bose systems with dipole-dipole
interaction. Moreover, for the anisotropic interaction the
situation is even more interesting. In particular, for a vertex with
two external $\rho$-lines $D_{\rho\rho}(K)$ we obtain
 \begin{eqnarray}\label{D_rhorho}
    D_{\rho\rho}(K\rightarrow 0)=
    \left(\frac{\partial \mu}{\partial \rho}\right)_{T}\left\{1+\epsilon(3k^2_z/k^2-1)\right\},
 \end{eqnarray}
where $\left(\partial \mu/\partial \rho\right)_T$ is the inverse
isothermal compressibility of the system and for convenience the
notation $\epsilon=g_{d}/\left(\partial \mu/\partial
\rho\right)_T$ is used. For the mechanical stability it is crucial
that $\epsilon<1$. A key moment in obtaining of this exact
identity is very similar to the original derivation of the
Hugenholtz-Pines relation \cite{Hugenholtz_Pines,Popov}. Taking
into account the fact that action (\ref{S}) is the Taylor series
expansion of the initial Eq.~(\ref{S_hydro}) near equilibrium
value of density it immediately follows that exact vertices in the
low-energy limit
\begin{eqnarray}\label{D_rho}
D_{\rho\rho\rho}(0,0,0)=\left(\partial^3 f/\partial
\rho^3\right)_T=\left(\partial^2 \mu/\partial
\rho^2\right)_T,\ldots
\end{eqnarray}
For the same reason we conclude that differentiation of every
exact vertex function with respect to $\rho$ gives the vertex with
one more zero-momentum $\rho(K)$ line.

To derive the second class of identities suppose that our system
is moving as a whole with velocity ${\bf v}$. In the
field-theoretic language it is equivalent to the following local
gauge transformation $\varphi(x)\rightarrow \varphi(x)-m{\bf
vr}/\hbar$ of the initial action (\ref{S_hydro}). Rotational
invariance in the $x-y$ plane ensures that the thermodynamic
quantities of a moving Bose system are functions of $ v_{\perp}^2$
and $v^2_z$ only, so for the free energy density we have
\begin{eqnarray}\label{f_v}
f_{\bf v}(\rho)=f(\rho)+\frac{m
    }{2}\left\{\rho^{\perp}_s v_{\perp}^2+\rho^{z}_sv_{z}^2 \right\}
    +o(v^2_z,v_{\perp}^2).
\end{eqnarray}
At zero temperature limit and in the absence of disorder
\cite{Pelster, Boudjemaa} the whole system is superfluid, i.e.,
$\rho^{\perp}_s=\rho^{z}_s=\rho$ and the last term in (\ref{f_v})
is equal zero identically, restoring Galilean invariant form of
the function $f_{\bf v}(\rho)$. Although we do not present the
explicit formula for gauge transformed action, but it is easy to
argue that the differential operator $\frac{\hbar}{im}{\bf k}
\frac{\partial}{\partial {\bf v}}$ acting on the bare vertex
function adds $\varphi(K)$ line with zero frequency and
vanishingly small ${\bf k}$. Of course, the same conclusions can
be drawn for the exact vertices. Thus despite of the complexity of
action (\ref{S}), the latter observation leads to (formally
treating the Matsubara frequency as a continuous variable)
\begin{eqnarray}\label{D_varphivarphi}
D_{\varphi\varphi}(K\rightarrow 0)&=&
\frac{\hbar^2}{m}\left\{\rho^{\perp}_sk^2_{\perp}+\rho^{z}_sk^2_z\right\},
\nonumber\\
D_{\varphi\rho}(K\rightarrow 0)&=&\omega_k,
\end{eqnarray}
and allows to shed light on the infrared structure of the
perturbation theory.

First of all, let us briefly discuss the long-length behavior of
third-order vertex functions. From the previous analysis we have
already found out that $D_{\rho\rho\rho}(K,Q,P)$ tends to constant
at small values of its arguments. Combining two above-mentioned
differentiation rules with respect to the density and velocity of
a moving Bose system we arrive with asymptotically exact result
\begin{eqnarray}\label{D_varphi}
&&D_{\varphi\varphi\rho}(K,Q|-K-Q)|_{K,Q \rightarrow
0}=\nonumber\\
&&-\frac{\hbar^2}{m}\left\{\left(\frac{\partial
\rho^{\perp}_s}{\partial \rho}\right)_T{\bf k}_{\perp}{\bf
q}_{\perp}+\left(\frac{\partial \rho^{z}_s}{\partial
\rho}\right)_Tk_{z}q_{z}\right\}.
\end{eqnarray}
The presence of even powers in the series expansion (\ref{f_v})
over velocity provides the following estimation for
$D_{\varphi\rho\rho}(K|Q,P)|_{K \rightarrow 0} \propto \omega_k
k^2_z$ when $k_z\gg k_{\perp}$ and $D_{\varphi\rho\rho}(K|Q,P)|_{K
\rightarrow 0} \propto \omega_k k^2_{\perp}$ in the opposite
limit. The main advantage of the hydrodynamic description is
absence of infrared divergences in perturbative calculations.
Therefore, it means that the behavior of the corresponding exact
correlation function is qualitatively reproduced even on the
one-loop level. The appropriate analysis leads to
$D_{\varphi\varphi\varphi}(K,Q,P)|_{K,Q,P \rightarrow 0}\propto
k_zq_z p^2_{\perp}\omega_p+{\rm perm.}$

The above results can be easily extended to higher-order vertices.
In general, the presence of two external $\varphi$-lines adjoins
to the appropriate vertex function
$D_{\varphi\varphi\ldots}(K,Q|\ldots)$ a factor ${\bf
k}_{\perp}{\bf q}_{\perp}$ (or $k_zq_z$); for the vertex with four
lines we find $D_{\varphi\varphi\varphi\varphi}(K,Q,P,S)\propto
({\bf k}_{\perp}{\bf q}_{\perp}p_z s_z+{\rm perm}.)$, and so on
for any vertex with even number of $\varphi$-lines at finite
temperatures. The low-energy result for functions with odd number
of external phase-field lines can be immediately obtained from the
limiting behavior of $D_{\varphi\varphi\varphi}(K,Q,P)$ and
$D_{\varphi\rho\rho}(K|Q,P)$.

Summarizing the results of this subsection, it should be noted
that in the long-length limit the quasiparticle spectrum exhibits
acoustic behavior $E_{\bf k}=\hbar k c_{\bf k}$ with anisotropic
sound velocity
\begin{eqnarray}
c^2_{\bf k}=\frac{1}{m}\left(\frac{\partial \mu}{\partial
\rho}\right)_{T}
\left\{\rho^{\perp}_s+(\rho^{z}_s-\rho^{\perp}_s)k^2_z/k^2\right\}\nonumber\\
\times\left\{1+\epsilon(3k^2_z/k^2-1)\right\},
\end{eqnarray}
and vertices given by Eqs.~(\ref{D_rho}), (\ref{D_varphi})
describe the effective interaction between phonons at low
temperatures.

\subsection{Anisotropic superfluid density}
Let us calculate the leading-order contribution to the so-called
normal densities $\rho^z_n$,  $\rho^{\perp}_n$ of a dipolar
superfluid at low temperatures. Nevertheless it is well-known
result \cite{Landau} for Bose systems with isotropic interaction
and the original Landau prescription is an easier one, but in
order to prove correctness of our approach we will perform these
calculations diagrammatically. The task is rather simple because
we have to find the long-length behavior of $\Pi_{\varphi\varphi}(K)$
at zero Matsubara frequency. In this limit the only relevant
vertex is $D_{\varphi\varphi\rho}(K,Q|P)$, so the result is given
by two diagrams (see Fig.~1).
\begin{figure}
\centerline{\includegraphics
[width=0.4\textwidth,clip,angle=0]{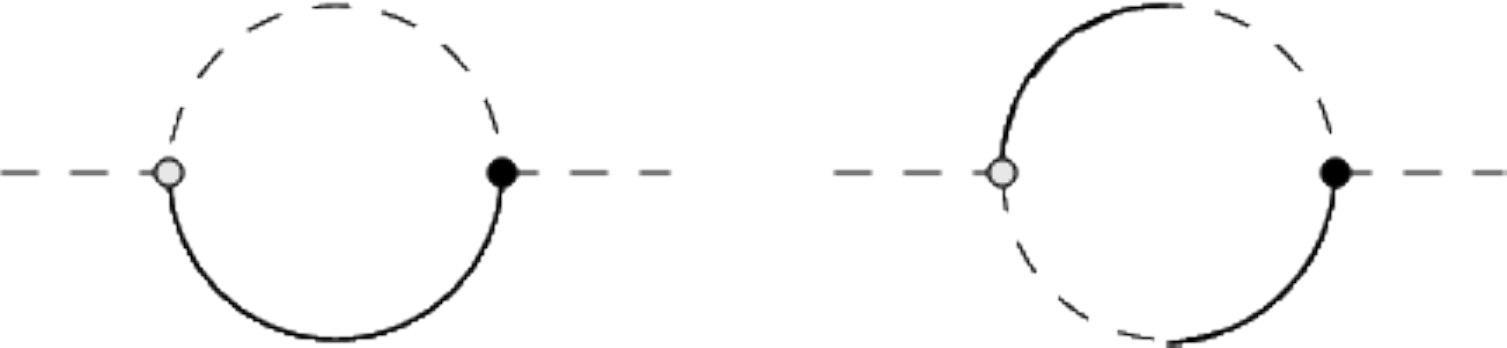}}
\caption{Diagrams contributing to the superfluid density. Dashed
and solid lines denote phase and density fields, respectively.
Light dot is $D^{(0)}_{\varphi\varphi\rho}$ and black dot stands
for the exact vertex function.}
\end{figure}
Moreover, at the low-temperature region we can neglect the
difference between superfluid densities and the total one, i.e.,
the following derivatives $\left(\partial \rho^{\perp}_s/\partial
\rho\right)_T$, $\left(\partial \rho^{z}_s/\partial \rho\right)_T$
in Eq.~(\ref{D_varphi}) may be readily replaced by unity
\begin{eqnarray}\label{Pi_varphi}
\Pi_{\varphi\varphi}(K\rightarrow 0)=\frac{1}{\beta
        V}\sum_{Q}\frac{\hbar^4}{m^2}({\bf kq})^2
        \left\{-\langle\varphi_Q\rho_{-Q}\rangle^2
        \right.\nonumber\\
\left.+\langle\varphi_Q\varphi_{-Q}\rangle\langle\rho_{Q}\rho_{-Q}\rangle\right\},
\end{eqnarray}
where it is enough to use asymptotic formulas (\ref{D_rhorho}),
(\ref{D_varphivarphi}) in equation (\ref{D_K}) for correlation
functions. After performing the Matsubara frequency summation and
calculation of the integrals in momentum space we obtain
\begin{eqnarray}\label{rho_n}
\rho^{\perp}_n&=&\frac{2\pi^2}{45}\frac{T^4}{mc^2(\hbar
c)^3}\frac{1}{(1-\epsilon)^2\sqrt{1+2\epsilon}}, \nonumber\\
\rho^{z}_n&=&\rho^{\perp}_n\frac{1-\epsilon}{1+2\epsilon},
\end{eqnarray}
where $mc^2=\rho\left(\partial \mu/\partial \rho\right)_{T}$. It
is worth noting that Eq.~(\ref{rho_n}) is exact in the
low-temperature limit and the attempt to perform such calculations
for the inverse isothermal susceptibility $\left(\partial
\mu/\partial \rho\right)_T$ is immediately met by considerable
difficulties. In fact, it is always easier to compute the
low-temperature behavior of all thermodynamic quantities taking
into account the phonon character of the excitation spectrum in the
long-length limit.

\subsection{Phonon damping}
The quasiparticle picture of the low-energy description provides
that the damping has to be small compared to  the energy of
elementary excitations. In general, the inverse life-time of
low-lying excitations written in terms of the imaginary
$\Im\Pi_{\rho\rho}(\omega,{\bf k})$,
$\Im\Pi_{\varphi\varphi}(\omega,{\bf k})$ parts and real $\Re
\Pi_{\varphi\rho}(\omega,{\bf k})$ part of appropriate
self-energies (after analytical continuation in the upper complex
half-plane $i\omega_k\rightarrow \omega+i0$) reads
\begin{eqnarray}\label{Gamma}
\Gamma_{\bf k}=\frac{\rho \hbar k}{2mc_{\bf
k}}\Im\Pi_{\rho\rho}(E_{\bf k}, {\bf k}) +\frac{mc_{\bf k}}{2\hbar
k\rho} \Im\Pi_{\varphi\varphi}(E_{\bf k}, {\bf k})\nonumber\\-\Re
\Pi_{\varphi\rho}(E_{\bf k}, {\bf k}).
\end{eqnarray}
Similarly to systems with isotropic interaction, the damping of
spectrum of a dipolar Bose gas is small in the long-length limit
and at very low temperatures. Actually, this fact enables to
calculate the leading-order contribution to the damping of phonon
mode exactly. Applying unitarity conditions at finite temperatures
\cite{Maleev} and using our estimations for the infrared structure
of the effective action from the previous section we evaluated the
imaginary parts of various analytically continued second-order
vertices (see Appendix for details).
\begin{figure} \centerline{\includegraphics
[width=0.4\textwidth,clip,angle=-0]{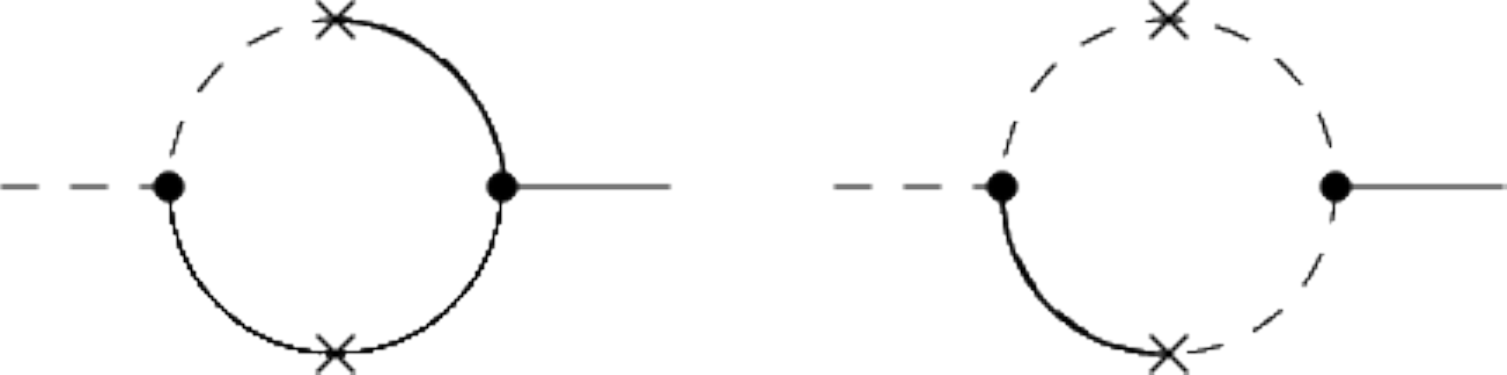}}
\caption{Diagrammatic representation of the leading-order
contribution to $\Re \Pi_{\varphi\rho}(\omega,{\bf k})$. Crosses
denote the spectral weights of the appropriate pair correlation
functions (see Ref.~\cite{Maleev} for details).}
\end{figure}
\begin{figure}
\centerline{\includegraphics
[width=0.4\textwidth,clip,angle=-0]{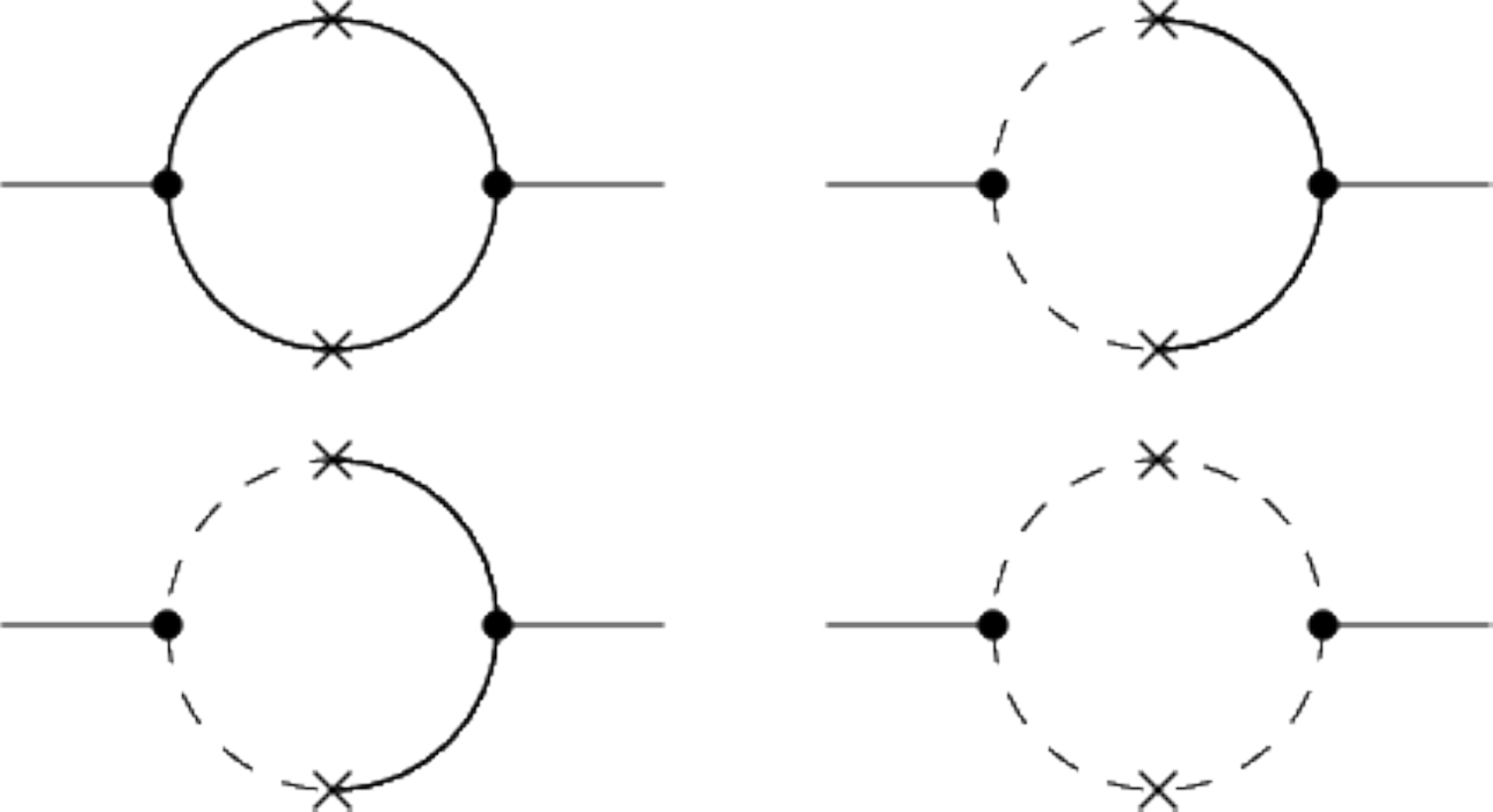}} \caption{Exact
low-energy asymptotics of $\Im \Pi_{\rho\rho}(\omega,{\bf k})$.}
\end{figure}

In zero-temperature limit the considerable contribution to the
damping is caused by the quasiparticle decay into two excitations
with lower energies. This mechanism which was originally proposed
in Ref.~\cite{Beliaev_2} and usually called the Beliaev damping is
also responsible for the decay of anisotropic phonons in a dipolar
Bose system
\begin{eqnarray}\label{Gamma_B}
\Gamma^B_{\bf k}=\frac{3}{640\pi}\frac{\hbar^2 k^5}{m\rho}
\frac{(1-\epsilon+3\epsilon
k^2_z/k^2)^{3/2}}{(1-\epsilon)\sqrt{1+2\epsilon}}\nonumber\\
\times\left\{1+\frac{\rho^2}{3c^2_{\bf k}}\frac{\partial}{\partial
\rho}\frac{c^2}{\rho}\right\}^2.
\end{eqnarray}
Of course, the processes when a given excitation decays into
three, four, and larger number of quasiparticles are also
possible, but their contribution to the damping is negligible in
the long-length limit $\hbar k/mc\ll 1$. Indeed, using a dimensional
analysis it is easy to argue that in the leading order the
three-phonon collisions give rise to the asymptotics of order
$k^9$ for the damping.

The situation changes at finite temperatures where the damping is
controlled by the Landau mechanism of quasiparticle decay. At the
low-temperature limit the inverse life-time $\Gamma_{\bf k}$ of
low-lying excitations is a linear function of wave-vector and taking
into account only binary quasiparticle collisions we obtain
\begin{eqnarray}\label{Gamma_L}
\Gamma^L_{\bf k}=\frac{3\pi^3}{40}\frac{\hbar k}{m
c}\frac{T^4}{(\hbar c)^3\rho} \frac{(1-\epsilon+3\epsilon
k^2_z/k^2)^{-1/2}}{(1-\epsilon)\sqrt{1+2\epsilon}}\nonumber\\
\times\left\{1+\frac{\rho^2}{3c^2_{\bf k}}\frac{\partial}{\partial
\rho}\frac{c^2}{\rho}\right\}^2.
\end{eqnarray}
It should be mentioned that the above formula is valid only in the
limit $k\ll mc/\hbar$ and for temperatures $T\ll mc^2$. In order
to explore the damping at higher temperatures and finite momenta
one has to take into account the contribution of all third-order
vertex functions \cite{Pastukhov}. Moreover in contrast to the
zero-temperature result, formula (\ref{Gamma_L}) is not an exact
low-temperature estimation of the decay rate because the $T^4$
terms are also present in the damping caused by three-phonon,
four-phonon, etc. scattering processes. Therefore, equation
(\ref{Gamma_L}) is accurate only in the dilute limit where it
generalizes the well-known result \cite{Hohenberg_Martin} on the
Bose systems with dipole-dipole interaction.
\section{Connection to a model with condensate}
\label{sec3}

The model with condensate is crucial for understanding properties
of interacting Bose particles. In fact, starting from the original
Bogoliubov theory \cite{Bogoliubov} very important results
concerning the field-theoretical description of these systems
\cite{Beliaev_1} low-energy behavior of the excitation spectrum
\cite{Hugenholtz_Pines,Gavoret_Nozieres} were obtained within this
approach. The purpose of this section is to show how the
long-length limit of the one-particle Green's function can be
reproduced using a hydrodynamic formulation. Therefore, our
objective is to study the properties of the following function
\begin{eqnarray}\label{psi*psi}
G(x-x')=-\langle
e^{i\varphi(x)}\sqrt{\rho(x)\rho(x')}e^{-i\varphi(x')}\rangle.
\end{eqnarray}
In general, these calculations cannot be done to the end, but for
our consideration it is enough to know the long-range behavior of
this function. At equal time-arguments ($\tau\rightarrow \tau'$)
(\ref{psi*psi}) up to a sign is equal to the one-body density
matrix $F({\bf r}-{\bf r}')$. The latter is the Fourier transform
of the particle number distribution in coordinate space and
$F(\infty)$ coincides with condensate density $\rho_{0}$. It is
intuitive that Eq.~(\ref{psi*psi}) suggests the condensate density
to have an exponential form. This very important feature of the
one-particle density matrix was mentioned for the first time in
Ref.~\cite{Penrose}. Due to non-commutativity of phase and density
operators (see Eq.~(\ref{psi})) the condensate fraction in the
hydrodynamic approach can be calculated as follows
\begin{eqnarray}\label{rho_0}
\sqrt{\rho_{0}}=\lim_{\tau'\rightarrow \tau-0}\langle
e^{i\varphi(x')}\sqrt{\rho(x)}\rangle|_{{\bf r}'={\bf r}},
\end{eqnarray}
or, equivalently $\lim_{\tau'\rightarrow \tau-0}\langle
\sqrt{\rho(x')}e^{-i\varphi(x)}\rangle|_{{\bf r}'={\bf r}}$. Of
course, the number of particles with zero momentum is a
model-dependent quantity, but the leading-order low-temperature
condensate depletion demonstrates universal power-law behavior
\cite{Ferrell}. Taking into account the above estimation for the
vertices one may show that
\begin{eqnarray}\label{rho_0_T}
\frac{\rho_{0}|_{T\rightarrow 0}}{\rho_{0}|_{T=0}}=1
-\frac{1}{V}\sum_{{\bf k} \neq 0}\left\{\frac{1}{
    \beta}\sum_{\omega_k}-\int^{\infty}_{-\infty}\frac{d\omega_k}{2\pi}\right\}
\nonumber\\
\times\langle\varphi_K\varphi_{-K}\rangle+\ldots.
\end{eqnarray}
At very low temperatures it is enough to substitute only the infrared
asymptotic of $\langle\varphi_K\varphi_{-K}\rangle$ in
Eq.~(\ref{rho_0_T}). After calculation of simple integrals we
arrive at the asymptotically exact result
\begin{eqnarray}\label{rho_0_as}
\frac{\rho_{0}|_{T\rightarrow 0}}{\rho_{0}|_{T=0}}=1-\frac{mT^2}{24\hbar^3\rho c}\frac{1}{
    \sqrt{3\epsilon}}\ln\left[\frac{\sqrt{1+2\epsilon}+\sqrt{3\epsilon}}{\sqrt{1+2\epsilon}-\sqrt{3\epsilon}}\right].
\end{eqnarray}

Making use of transformation to the four-momentum space for the
one-particle Green's function we find
\begin{eqnarray}\label{G_P}
\mathcal{G}(P)=\int dx\, e^{-iPx}\left\{\rho_{0}+G(x)\right\},
\end{eqnarray}
and taking into account equations (\ref{psi*psi}) and
(\ref{rho_0}) we have
\begin{eqnarray}\label{G_P_0}
\mathcal{G}(P\rightarrow
0)=-\rho_{0}\langle\varphi_P\varphi_{-P}\rangle.
\end{eqnarray}
Actually, the above equation generalizes the well-known
Gavoret-Nozi\`eres result \cite{Gavoret_Nozieres}, which was also
obtained in Ref.~\cite{Nepomnyashchy} with taking into
consideration correct infrared properties of self-energies. More
recently the low-energy behavior of the one-particle Green's
function was analyzed using renormalization group techniques
\cite{Pistolesi, Dupuis} and two-fluid hydrodynamics
\cite{Arahata}. At finite temperatures the zero-frequency limit of the
formula (\ref{G_P_0}) is the extension of the Josephson result
\cite{Josephson} on systems with anisotropic interaction. It is
also clear that due to the presence of an external field that aligns
all dipoles along one direction, equation (\ref{G_P}) possesses an
anisotropic dependence on the wave-vector, which for the
two-dimensional system leads to very intriguing non-diagonal
long-range behavior of the one-body density matrix $F({\bf r})$.

Finally, the same analysis can be easily applied to the anomalous
one-particle Green's function. Without going into details we note
that the infrared asymptotic behavior of this function up to a
sign coincides with (\ref{G_P_0}).
\section{Propagation of sound modes}
\label{sec4}
Now we briefly discuss peculiarities of macroscopic
two-fluid hydrodynamics in Bose systems with anisotropic
interaction. In particular we are interested in describing of the
small-amplitude oscillations of various physical quantities near
their equilibrium values. Thus, local thermodynamic equilibrium
is assumed. Moreover, it is also supposed that all observables are
smooth functions of spatial coordinates ${\bf r}$ and time $t$
although we will not write this explicit dependence. Our further
consideration with minor changes is adopted from
\cite{Pitaevskii_Stringari}. The local density should satisfy the
continuity equation
\begin{eqnarray}\label{rho_cont}
\partial_t\rho+\textrm{div}\, {\bf j}=0,
\end{eqnarray}
where the density current consists of two terms ${\bf j}={\bf
j}_s+{\bf j}_n$. The first one is the current of the superfluid
component ${\bf j}_s=(\rho^{\perp}_s{\bf v}^{\perp}_s, \rho^z_s
v^z_s)$, which correctly reproduces the anisotropic behavior of the
system, and the second ${\bf j}_n=(\rho^{\perp}_n{\bf
v}^{\perp}_n, \rho^z_n v^z_n)$ corresponds to the normal component
of the dipolar Bose system. The evolution of the superfluid velocity
is governed by the following equation
\begin{eqnarray}\label{v_s}
\partial_t{\bf v}_s+\frac{1}{m}\nabla \mu_d=0,
\end{eqnarray}
where the chemical potential of a slightly non-uniform system
contains the additional term $\mu_d({\bf r},t)=\mu(\rho({\bf
r},t))+\int d {\bf r}'\Phi_{d}({\bf r}-{\bf r}')\rho({\bf r}',t)$.
The same concerns the local pressure $p_d$ in the equation for the
density current
\begin{eqnarray}\label{current}
\partial_t\,{\bf j}+\frac{1}{m}\nabla p_d=0.
\end{eqnarray}
At last we have to write down the equation for the entropy per
unit volume $s$ which together with thermodynamic identity
\begin{eqnarray}\label{therm_eq}
\rho d\mu_d=-sdT+dp_d,
\end{eqnarray}
close the system of equations for sound propagation. From the
general arguments it is clear that the entropy $s$ is proportional
to the number of quasiparticles $n_T=\frac{1}{V}\sum_{{\bf q}\neq
0}(e^{\beta E_{\bf q}}-1)^{-1}$ in a unit volume of the system. On
the other hand, by using kinetic equation for the quasiparticle
distribution function \cite{Griffin_Nikuni_Zaremba} (in the frame,
where superfluid component is at rest ${\bf v}_s=0$) it is easy to
show that $n_T$ should satisfy the continuity equation, which can
be immediately rewritten for the entropy density
\begin{eqnarray}\label{s_cont}
\partial_ts+\textrm{div}(s {\bf v}_n)=0.
\end{eqnarray}
Note that instead of the density current (\ref{current}) the entropy
transport is isotropic. The further derivation of sound velocities
is almost standard. Combining equations (\ref{rho_cont}) and
(\ref{current}) one gets
\begin{eqnarray}\label{1_hydro}
\partial^2_t\rho=\frac{1}{m}\Delta p_d.
\end{eqnarray}
Introducing the entropy per particle $\tilde{s}=s/\rho$ and making use
of Eqs.~(\ref{v_s})-(\ref{s_cont}) we finally obtain
\begin{eqnarray}\label{2_hydro}
\partial^2_t\tilde{s}=\frac{\tilde{s}^2}{m}
\left\{(\rho^{\perp}_s/\rho^{\perp}_n)\Delta_{\perp}+(\rho^{z}_s/\rho^{z}_n)\partial^2_z
\right\}T.
\end{eqnarray}
Taking the local density and temperature as independent variables
and after substitution of the plane-wave solution $\rho, T \propto
e^{i({\bf kr}-ukt)}$ the obtained secular equation that determines
sound velocities reads
\begin{widetext}
\begin{eqnarray}\label{sec_eq}
\left\{u^2-\frac{1}{m}\left(\frac{\partial p}{\partial
\rho}\right)_T[1+\epsilon(3\cos^2\theta-1)]\right\}\left\{\frac{\tilde{c}_{\rho}}{T}u^2-
\frac{\tilde{s}^2}{m}\left[\frac{\rho^{\perp}_s}{\rho^{\perp}_n}\sin^2\theta+\frac{\rho^{z}_s}{\rho^{z}_n}\cos^2\theta\right]\right\}
-\frac{u^2}{m}\left[\frac{1}{\rho}\left(\frac{\partial p}{\partial
T}\right)_{\rho}\right]^2=0,
\end{eqnarray}
\end{widetext}
where $\tilde{c}_{\rho}=T(\partial \tilde{s}/\partial T)_{\rho}$
is the specific heat per particle at constant volume and $p$ is
the pressure of a uniform dipolar Bose system. Here we also
introduced the angle $\theta$ between the direction of sound wave
propagation and $z$ axis. At low temperatures the velocity of the
density perturbations (first sound) is
\begin{eqnarray}
u^2_1=c^2\{1+\epsilon(3\cos^2\theta-1)\},
\end{eqnarray}
and taking into account Eq.~(\ref{rho_n}) as well as limiting
behavior of various thermodynamic quantities for the velocity of
temperature waves (second sound) we obtain
\begin{eqnarray}
u^2_2=u^2_1/3.
\end{eqnarray}
In the normal phase Eq.~(\ref{sec_eq}) has a unique solution
\begin{eqnarray}
u^2=\frac{1}{m}\left(\frac{\partial p}{\partial
\rho}\right)_{\tilde{s}}+\frac{\rho g_{d}}{m}(3\cos^2\theta-1)
\end{eqnarray}
that recovers the usual velocity of adiabatic pressure waves in the
fully polarized dipolar system. In the magnetic field of finite
magnitude the above hydrodynamic equations should be complemented
with the equation for magnetization \cite{Andreev_Kuzmenkov}.
\section{Conclusions}
\label{sec4}
In conclusion, we have studied the properties of Bose
systems with dipole-dipole interaction. Particularly, by means of
hydrodynamic approach we analyzed the exact infrared structure of
low-energy action that describes density and phase fluctuations.
Within these results we calculate the spectrum and damping of
collective modes in terms of macroscopic quantities of a dipolar
Bose system, find out the long-length behavior of the one-particle
Green's functions and perform calculations of low-temperature
corrections to the anisotropic superfluid density. The
leading-order temperature dependence of the condensate fraction is
also obtained. Additionally we considered the linearized
macroscopic hydrodynamics of dipolar superfluids and explored
peculiarities of first and second sounds. These findings clearly
demonstrate the impact of dipole-dipole interaction on the
anisotropic low-energy properties of the system.
\begin{center}
{\bf Acknowledgements}
\end{center}
We thank Prof.~I.~Vakarchuk and Dr.~A.~Rovenchak for stimulating
discussions. This work was partly supported by Project FF-30F
(No.~0116U001539) from the Ministry of Education and Science of
Ukraine.
\section{Appendix}
\setcounter{equation}{0} In this section we present the explicit
expressions for the imaginary parts of the exact second-order
vertices in the low-length limit. The leading-order contribution
is determined by all diagrams with only two vertices, namely
$D_{\varphi\varphi\rho}(K,Q|P)$ and $D_{\rho\rho\rho}(K,Q,P)$.
Feynman diagrams contributing to $\Re
\Pi_{\varphi\rho}(\omega,{\bf k})$ and $\Im
\Pi_{\rho\rho}(\omega,{\bf k})$ are given in Fig.~2 and Fig.~3,
respectively. In the same manner the low-energy behavior of $\Im
\Pi_{\varphi\varphi}(\omega,{\bf k})$ can be easily obtained from
Fig.~1. For convenience terms contributing to the Beliaev damping
\begin{widetext}
\begin{eqnarray}
\Im \Pi^B_{\varphi\varphi}(\omega,{\bf
k})=\frac{\pi}{4V}\sum_{{\bf q}\neq
    0} \frac{\hbar^4{\bf kq}}{m^2}\left[{\bf kq}\frac{c_{\bf q}}{q}
\frac{|{\bf q}+{\bf k}|}{c_{\bf q+k}}-{\bf k}({\bf q}+{\bf
k})\right] \delta(E_{{\bf q}+{\bf k}}+E_{\bf q}-\omega),
\end{eqnarray}
\begin{eqnarray}
\Im \Pi^B_{\rho\rho}(\omega,{\bf k})=\frac{\pi}{8V}\sum_{{\bf
q}\neq
    0} \frac{\hbar^2q|{\bf q}+{\bf k}|}{\rho^2c_{\bf q}c_{\bf
        q+k}}\left[\frac{{\bf q}({\bf q}+{\bf k})}{q|{\bf q}+{\bf
        k}|}c_{\bf q}c_{{\bf q}+{\bf k}}-\rho^2\frac{\partial
    }{\partial \rho}\frac{c^2}{\rho}\right]^2\delta(E_{{\bf q}+{\bf k}}+E_{\bf q}-\omega),
\end{eqnarray}
\begin{eqnarray}
\Re \Pi^B_{\varphi\rho}(\omega,{\bf k})= -\frac{\pi}{4V}\sum_{{\bf
        q}\neq 0} \frac{\hbar^3{\bf kq}}{m\rho}\left[\frac{{\bf q}({\bf
        q}+{\bf k})}{q}c_{\bf q}
-\frac{|{\bf q}+{\bf k}|}{c_{\bf q+k}}\rho^2\frac{\partial
    }{\partial \rho}\frac{c^2}{\rho}\right]\delta(E_{{\bf q}+{\bf k}}+E_{\bf q}-\omega),
\end{eqnarray}
and the Landau damping
\begin{eqnarray}
\Im \Pi^L_{\varphi\varphi}(\omega,{\bf k})=
-\omega\frac{\pi}{V}\sum_{{\bf q}\neq 0} \left[\frac{\hbar^2{\bf
kq}}{m}\right]^2\left\{\frac{\partial}{\partial E_{\bf q}}n(\beta
E_{\bf q})\right\}\delta(E_{{\bf q}+{\bf k}}-E_{\bf q}-\omega),
\end{eqnarray}
\begin{eqnarray}
\Im \Pi^L_{\rho\rho}(\omega,{\bf k})=
-\omega\frac{\pi}{4V}\sum_{{\bf q}\neq 0}\left(\frac{E_{\bf
q}}{\rho}\right)^2 \left[1+\frac{\rho^2}{c^2_{\bf
q}}\frac{\partial }{\partial
\rho}\frac{c^2}{\rho}\right]^2\left\{\frac{\partial}{\partial
E_{\bf q}}n(\beta E_{\bf q})\right\}\delta(E_{{\bf q}+{\bf
k}}-E_{\bf q}-\omega),
\end{eqnarray}
\begin{eqnarray}
\Re \Pi^L_{\varphi\rho}(\omega,{\bf k})=
\omega\frac{\pi}{2V}\sum_{{\bf q}\neq 0}\frac{\hbar^2 {\bf
kq}}{m}\frac{E_{\bf q}}{\rho}\left[1+\frac{\rho^2}{c^2_{\bf
q}}\frac{\partial }{\partial
\rho}\frac{c^2}{\rho}\right]\left\{\frac{\partial}{\partial E_{\bf
q}}n(\beta E_{\bf q})\right\}\delta(E_{{\bf q}+{\bf k}}-E_{\bf
q}-\omega),
\end{eqnarray}
are written separately.
\end{widetext}

\end{document}